\documentclass[prl,aps,floats,floatfix,twocolumn,showpacs,letterpaper]{revtex4}

\usepackage[dvips]{graphicx}
\usepackage{amsmath} 
\usepackage{psfrag}
\usepackage{longtable}
\usepackage{dcolumn,epsfig}

\def\laq{\raise 0.4ex\hbox{$<$}\kern -0.8em\lower 0.62ex\hbox{$\sim$}}
\def\gaq{\raise 0.4ex\hbox{$>$}\kern -0.7em\lower 0.62ex\hbox{$\sim$}}

\newcommand{\beq}{\begin{equation}}
\newcommand{\eeq}{\end{equation}}
\newcommand{\bea}{\begin{eqnarray}} 
\newcommand{\eea}{\end{eqnarray}}
\newcommand{\ba}{\begin{array}}
\newcommand{\ea}{\end{array}}

\newcommand{\comment}[1]{}

\newlength{\sizeonefig}
\newlength{\sizetwofig}
\newlength{\sizeonefigb}
\newlength{\sizetwofigb}
\begin{document}
\title{Interferometers for Displacement-Noise-Free Gravitational-Wave Detection}

\author{Yanbei Chen, Archana Pai, Kentaro Somiya }
\affiliation{
Max-Planck Institut f\"ur Gravitationsphysik, Am M\"uhlenberg 1, 14476 Potsdam, Germany
}
\author{
 Seiji Kawamura, Shuichi Sato}
 \affiliation{TAMA project, National Astronomical Observatory of Japan, 
2-21-1 Osawa, Mitaka, Tokyo 181-8588, Japan}
\author{Keiko Kokeyama}
\affiliation{
Ochanomizu University, 2-1-1, Otsuka, Bunkyo-ku, Tokyo, 112-8610 Japan}
\author{Robert L. Ward} 
\affiliation{
LIGO Project 18-34, California Institute of Technology, Pasadena, California 91125, USA}


\date{March 13, 2005}
\begin{abstract} 
We propose a class of displacement- and laser-noise free gravitational-wave-interferometer configurations, which does not sense non-geodesic mirror motions and laser noises, but provides non-vanishing gravitational-wave signal.  Our interferometer consists of 4 mirrors and 2 beamsplitters, which form 4 Mach-Zehnder interferometers. By contrast to previous works, no composite mirrors are required. Each mirror in our configuration is sensed redundantly, by
at least two pairs of incident and reflected beams. Displacement- and laser-noise free detection is achieved  when output signals from these 4 interferometers are combined appropriately.  Our 3-dimensional interferometer configuration has a low-frequency response proportional to $f^2$, which is better than the $f^3$ achievable by previous 2-dimensional configurations.
\end{abstract}
\maketitle

It was recently demonstrated theoretically that gravitational-wave (GW) detection does not require freely falling test masses, because non-geodesic test-mass motion 
affect travel times of pulses only when they arrive and leave the test masses, while the effect of GWs are distributed~\cite{KC1}. This idea was further explored in Ref.~\cite{KC2}, which shows that once the number $\mathcal{N}$ of test masses is large enough, the number of light-pulse-travel-time measurement channels between test masses [$O(\mathcal{N}^2)$] will  exceed the total number of clock- and displacement-noise channels [$O(\mathcal{N})$], and there must exist clock- and displacement-noise-free channels.  Ref.~\cite{KC2} also showed that interferometers can be combined to realize displacement- and laser-noise free GW detection.  
As argued there, when lasers are used as part of the detection strategy, motions of the laser devices cause Doppler shift to the laser frequencies, and are indistinguishable from laser noises.  Therefore, displacement-noise-free detection, strictly speaking, requires the cancelation of laser noise. Henceforth, we shall   use the term Displacement-noise-free Interferometry (DFI) to describe displacement- and laser-noise free interferometer configurations. 

Refs.~\cite{KC1} and \cite{KC2} study DFI by calculating pulse time delays between emitters and receivers, which are fixed on point test masses. This approach, although
mathematically simpler and in principle applicable to laser interferometry, does not provide {\it practical} interferometer configurations right away.  In particular, interferometer configurations constructed so far require {\it composite mirrors,} namely mirrors with multiple reflective surfaces. Apart from being experimentally challenging, the use of composite mirrors gives rise to the fundamental difficulty that thermal fluctuations of relative positions between the multiple reflective surfaces are not canceled.  In addition, so far only 2-dimensional configurations have been explored, for which it can be proved that sensitivity to GWs can be no better than $\sim f^3$ in low frequencies~\cite{KC4}.

\begin{figure}
\includegraphics[width=0.3\textwidth]{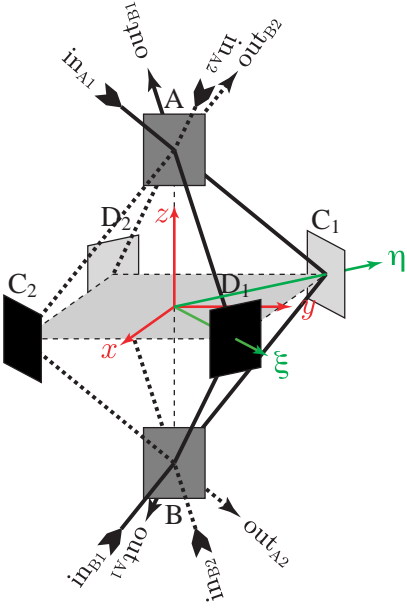}
\caption{The 3-D octahedron configuration, which consists of 4 Mach-Zehnder interferometers; $A_1:\, A_tC_1B_r-A_rD_1B_t$, $B_1:\, B_tC_1A_r-B_rD_1A_t$ (shown as solid lines) and $A_2:\,  A_tD_2 B_r- A_rC_2B_t$, $B_2:\,  B_t D_2 A_r-B_rC_2A_t$ (shown as dashed lines). The subscripts $r,t$ indicate reflection and transmission.  \label{3dMZ}}
\end{figure}

In this paper, we propose a class of 2-D and 3-D interferometers that implement DFI without using composite mirrors.  Although these configurations were initially discovered using linear algebraic manipulations within the time-delay formulation in Ref.~\cite{KC2}, they turn out to have very simple physical interpretations.  First of all, we use Mach-Zehnder interferometry, so that laser noises can be canceled right in the beginning. Moreover, each mirror participates in at least two Mach-Zehnder interferometers, and thus has its location sensed redundantly. Finally, by combining the Mach-Zehnder output signals, we are able to cancel amongst the redundant displacement information, leaving non-vanishing response to GWs. In particular, we will show that our 3-D configuration has $\sim f^2$ sensitivity in low frequencies, which is the best one can achieve with DFI~\cite{KC4}.  The 2-D configuration, which has $\sim f^3$ response in low frequencies, is proposed mainly for the purpose of initial experimental tests.

{\it GW response of a plane electromagnetic (EM) wave.} For self-containedness, we provide a brief derivation of the GW-induced phase shift of light. A weak plane GW on Minkowski background can be described with a metric 
\begin{equation}
g_{\mu\nu} = \eta_{\mu\nu} + h_{\mu\nu}\,,
\end{equation}
in the Cartesian coordinate system, $x^\mu=(t,\mathbf{x})$, with $t$ the time coordinate, $\mathbf{x}$ the spatial coordinates, and  $\eta_{\mu\nu} = {\rm diag}(-1,1,1,1)$. In the  Transverse-Traceless (TT) gauge, $h_{\mu \nu}$ only has spatial components:
\begin{eqnarray}
\label{hTT}
\mathbf{h}^{\rm TT}(t,\mathbf{x})\!\!& = &\!\! h_+ (t- \mathbf{e}_Z \cdot \mathbf{x}/c) \left[\mathbf{e}_X \otimes \mathbf{e}_X
-\mathbf{e}_Y \otimes \mathbf{e}_Y\right] \nonumber \\
\!\! & + & \!\! h_\times (t- \mathbf{e}_Z \cdot \mathbf{x}/c) \left[\mathbf{e}_X \otimes \mathbf{e}_Y
+ \mathbf{e}_Y \otimes \mathbf{e}_X\right].\quad
\end{eqnarray}
Here $(\mathbf{e}_X,\mathbf{e}_Y,\mathbf{e}_Z)$ is a spatial orthonormal set, with $\mathbf{e}_Z$ the wave propagation direction.  We approximate the EM field as a scalar wave, with amplitude 
\begin{equation}
\Phi(x^\mu) = \Phi^{(0)}(x^\mu) \left[1+i \phi^{\rm gw}(x^\mu)\right]\,,
\end{equation}
with $\Phi^{(0)}(x^\mu) = A \exp(i k_\mu x^\mu)$ the 0-th order EM wave when there is no GW ($A$ is constant), and $\phi^{\rm gw}(x^\mu)$ the additional  phase shift caused by the GW.  The EM wave equation $g^{\mu\nu}  \Phi_{;\mu\nu} =0$, expanded to leading order in $h_{\mu\nu}$ and $\phi^{\rm gw}$, can be written as
\begin{equation}
\eta^{\mu\nu}  \Phi_{,\mu\nu} = - h^{\mu\nu}  \Phi^{(0)}_{,\mu\nu}\,,
\end{equation}
where the Lorenz gauge condition ${h^{\mu\nu}}_{,\nu}=0$ has been used. Because $\phi^{\rm gw}$ is slowly varying compared to $\Phi^{(0)}$, we can ignore terms like $\phi^{\rm gw}_{,\mu\nu}$ on the left-hand side, and obtain
\begin{equation}
k^\nu  \phi^{\rm gw}_{,\nu} = h_{\mu\nu} k^{\mu} k^{\nu}/2\,,
\end{equation}
which accumulates along the path of the light ray in Minkowski spacetime.  In particular, if the Minkowski ray starts from $(t_0,\mathbf{x}_0)$ and ends at $(t,\mathbf{x})$, with $|\mathbf{x}- \mathbf{x}_0|= |t-t_0|=l$ and $\mathbf{N} \equiv (\mathbf{x}- \mathbf{x}_0)/l$, then the GW-induced phase shift is
\begin{eqnarray}
\label{gwresp}
&&\phi^{\rm gw}(t_0,\mathbf{x}_0;\mathbf{x})  \nonumber \\
&=& {\omega l}/(2c)\int_0^1 d\zeta  h_{ij}^{\rm TT}(t_0 + l \zeta,\mathbf{x}_0 + \mathbf{N}l \zeta)N_i N_j \,.
\end{eqnarray}

{\it 3-D Configuration.}  We now discuss our 3-D configuration,
shown in Fig.~\ref{3dMZ}.  The mirrors are located on the 8 vertices
of a {\it regular octahedron,} with edge length $2L$. All light
rays in our interferometer will be propagating along the edges of the
octahedron. A Cartesian coordinate system is attached
to the octahedron, with the origin coinciding with its center, $z$
axis coinciding with its $B$-$A$ axis, $x$ axis parallel to the
$C_1$-$D_1$ ($D_2$-$C_2$) direction, and $y$ axis parallel to the
$C_2$-$D_1$ ($D_2$-$C_1$) direction.  [We have also defined $\xi$ and
$\eta$ directions, as shown in the figure.] A $50$-$50$ beamsplitter each
is located on the vertices $A$ and $B$, with normal directions parallel to 
the $x$ axis. The four perfectly reflective mirrors at $C_{1,2}$ and $D_{1,2}$
are such oriented that light rays from $A$ will be reflected directly
to $B$.  We assume all perfect mirrors to have amplitude
reflectivity $r=1$, and both 50-50 beamsplitters to have $-1/\sqrt{2}$ amplitude 
reflectivity for light incident from the $+x$ side (i.e., traveling
toward $-x$ direction), and $+1/\sqrt{2}$ amplitude reflectivity for light incident from the $-x$ side; 
the edge
length is assumed to be an
integer multiple of the optical wavelength, at the zero point of the device (i.e., in absense of laser noise, non-geodesic mirror motion, and GW). 

In this mirror set-up, we construct {\it four Mach-Zehnder interferometers,}
$A_1$, $B_1$ (with light paths in solid lines), $A_2$ and $B_2$ (with
light paths in dashed lines), with input and output ports indicated in
Fig.~\ref{3dMZ}.  At the zero point, the ports
$\mbox{out}_{A_1}$, $\mbox{out}_{A_2}$, $\mbox{out}_{B_1}$, and
$\mbox{out}_{B_2}$ are all dark, while each input port is also the bright port for another  interferometer. During operations, for each Mach-Zender, $I=A_1,A_2,B_1,B_2$, if $\phi^{(t)}_I$ and $\phi^{(r)}_I$ represent the additional phase shifts gained by the beams transmitted and reflected from its first beamsplitter, respectively, then the output optical amplitude is proportional to
\begin{equation}
\label{singlemz}
e^{ i \phi^{(t)}_I } - e^{ i \phi^{(r)}_I }  \propto \phi^{(t)}_I -  \phi^{(r)}_I \equiv \delta \phi_I  \,.
\end{equation}
For interferometers $A_1$ and $A_2$, the ``first beamsplitter'' means $A$, while for $B_1$ and $B_2$, it means $B$. In Eq.~\eqref{singlemz}, we always have the minus sign in front of $ e^{ i \phi^{(r)}_I }$, because the lasers always incident from the $+x$ side of the beamsplitters, and hence the first reflection always encounters a $-1/\sqrt{2}$ amplitude reflectivity.

\begin{figure}
\centerline{\epsfig{file=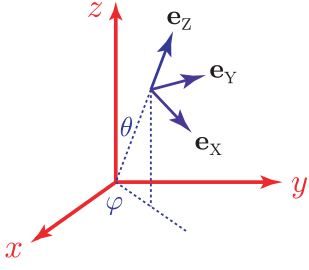, width=0.225\textwidth}}
\caption{The orthonormal system $(\mathbf{e}_X,\mathbf{e}_Y,\mathbf{e}_Z)$  used to describe a generic plane GW; $\mathbf{e}_Z$ is the propagation direction.  \label{waveframe}}
\end{figure}

Physically, the additional phase shifts can arise from laser noise, displacement noise, and GWs. Because we only consider  linear order in GWs and the noises, we can first include only effects of laser and displacement noises, construct a combination from the outputs of the four Mach-Zehnder interferometers that is free from these noises, and then calculate its response to GWs.  For dark-port detection, each Mach-Zehnder interferometer is already free from laser noise; we only need to evaluate their displacement sensitivities. For a mirror with normal direction $\mathbf{n}$ and incident wavevector $\mathbf{k}$, the phase shift gained by the reflected light when the mirror moves spatially by $\mathbf{\delta x}$ is $2 (\mathbf{n} \cdot \mathbf{k} )(\mathbf{n} \cdot \mathbf{\delta x})$.  For $A_1$ and $B_1$ interferometers, we have
\begin{eqnarray}
\phi^{(t)}_{A_1}(t) \!\!&=& \!\! {\sqrt{2}\omega}/{c} \left[\eta_{C_1}(t-2L /c) +x_B(t)\right]\,,  \\
\phi^{(r)}_{A_1}(t)\!\! &=&\!\!  {\sqrt{2}\omega}/{c} \left[\xi_{D_1}(t-2L /c) -x_A(t-4L /c)\right]\,, \\
\phi^{(t)}_{B_1}(t)\!\! &=& \!\! {\sqrt{2}\omega}/{c} \left[\eta_{C_1}(t-2L /c) +x_A(t)\right]\,,  \\
\phi^{(r)}_{B_1}(t)\!\! &=& \!\! {\sqrt{2}\omega}/{c} \left[\xi_{D_1}(t-2L /c) -x_B(t-4L /c)\right]\,.
\end{eqnarray}
Here, $\eta_{C_1}$ denotes the motion of $C_1$ along the $\eta$ axis and so on.
Thus we have
\begin{eqnarray}
&&\delta\phi_{A_1}-\delta\phi_{B_1} =\left[\phi_{A_1}^{(t)}-\phi_{A_1}^{(r)}\right]-\left[\phi_{B_1}^{(t)}-\phi_{B_1}^{(r)}\right] \nonumber \\
 \!\!&=&\!\! {\sqrt{2}\omega}/{c} 
 \big[x_B(t)-x_A(t)  \nonumber \\ 
&&\quad \quad \; - x_B(t-4L /c)+x_A(t-4L /c)\big]. \;\quad
\end{eqnarray}
Here, we have denoted with $\omega$ the optical frequency and $c$ the speed of light. Note that motions of $C_1$ and $D_1$ are already canceled in this subtraction, because the two Mach-Zehnders sense their motions equally, due to the fact that $|AD_1|=|BD_1|=|AC_1|=|BC_1|$. Similarly,  we have a combination of the other two Mach-Zehnders: 
\begin{eqnarray}
&&\delta\phi_{A_2}-\delta\phi_{B_2}  \nonumber \\
 \!\!&=&\!\! {\sqrt{2}\omega/c} 
 \big[x_B(t)-x_A(t)  \nonumber \\ 
&&\quad \;   \; - x_B(t-4L /c)+x_A(t-4L /c)\big]. \;\quad
\end{eqnarray}
As a consequence, the total combination 
\begin{equation}
\phi_{\rm DFI} \equiv \left[\delta\phi_{A_1}-\delta\phi_{B_1}\right] -\left[\delta\phi_{A_2}- \delta\phi_{B_2}\right].
\end{equation}
is free from any displacement noise. This is also anticipated, because it is obvious that $A_1$ and $A_2$ sense the beamsplitters in the same way, and so do  $B_1$ and $B_2$.

\begin{figure}[t]
\psfrag{xx}[t][]{$2\Omega L/(\pi c)$}
\psfrag{yy}[b][]{$\left[\phi^{\rm gw}/(\omega L h)\right]_{\rm rms}$}
\centerline{\epsfig{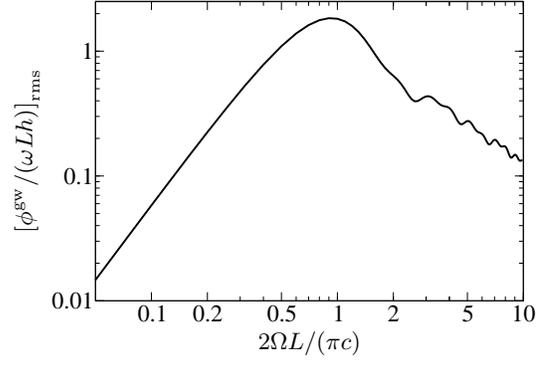}}
\caption{\label{fig:resp} Root-mean-square transfer function of the 3-D, 4-Mach-Zehnder configuration.}
\end{figure}

We now  calculate the response of $\phi_{\rm DFI}$ to GWs. For a particular case, with a plane GW coming directly along the $z$ axis [i.e., $\mathbf{e}_Z = \mathbf{e}_z$, Cf.~Eq.~\eqref{hTT}]  and
\begin{equation}
\mathbf{h}^{\rm TT}(t,\mathbf{x}) = h (t-z/c)\left[\mathbf{e}_\xi \otimes \mathbf{e}_\xi - \mathbf{e}_\eta \otimes \mathbf{e}_\eta\right]\,,
\end{equation}
it is easy to argue based on octahedron's symmetry, and the way we combine the signals, that beams in all four {\it branches}, i.e., those involving $C_1$, $D_1$, $C_2$ and $D_2$ respectively, will give equal GW contributions to the final combination.  For one particular branch, involving $D_1$, we calculate the GW response, using Eq.~\eqref{gwresp}:
\begin{eqnarray}
\frac{\phi^{\rm gw}_{\rm DFI}}{\omega L/c} 
= \frac{1}{2}\int_0^1 d\zeta \!\!\!&\big\{&\!\!\! h [t+(2L\zeta - \sqrt{2} L (1-\zeta))/c] \nonumber \\
\!\!\!&+&\!\! \!  h[t+({2L(1+\zeta) + \sqrt{2} L \zeta})/{c}] \nonumber \\
\!\!\!&-&\!\!\!  h [t+({2L\zeta + \sqrt{2} L(1-\zeta)})/{c}] \nonumber \\
\!\!\!&-& \!\!\! h [t+({2L(1+\zeta) - \sqrt{2}L\zeta})/{c}]\big\}.\quad
\end{eqnarray}
In the frequency domain, we have
\begin{eqnarray}
\tilde{\phi}^{\rm gw}_{\rm DFI}&=& {i\omega {\tilde h} e^{-i\sqrt{2}\Omega L/c} }/({4\Omega}) \nonumber \\
&&\Big[(2-\sqrt{2})[1- e^{(4+2\sqrt{2})i\Omega L/c}] \nonumber \\
&&+(2+\sqrt{2})[e^{4i\Omega L/c}-e^{2\sqrt{2}i\Omega L/c}]\Big],
\end{eqnarray}
where $\tilde \phi^{\rm gw}_{\rm DFI}$ and $\tilde h$ are Fourier transforms of $\phi^{\rm gw}_{\rm DFI}$ and $h$.   This already shows a non-vanishing response. 

For GWs with generic propagation directions and polarizations, we use the following notation (Cf.~Eq.~\eqref{hTT}
 and Fig.~\ref{waveframe}),
\begin{eqnarray}
\mathbf{e}_X &=&
 \mathbf{e}_x  \cos\theta\cos\varphi+    \mathbf{e}_xy\cos\theta\sin\varphi -  \mathbf{e}_z  \sin\theta\,,\\
\mathbf{e}_Y &=&  - \mathbf{e}_x  \sin\varphi +  \mathbf{e}_y  \cos\varphi \,, \\
\mathbf{e}_Z &=&  \mathbf{e}_x  \cos\varphi\sin\theta +  \mathbf{e}_y  \sin\theta\sin\varphi +  \mathbf{e}_z \cos\theta\,.
\end{eqnarray}
In low frequencies, the GW response of the DFI combination is $\sim f^2$, with ~\footnote{This angular response, although looks similar to that of a Michelson interferometer lying on the $x$-$y$ plane, in fact cannot be put into that form, regardless of the orientation of the Michelson. In particular, the response of a Michelson with arms along $x$ and $y$ is 
$ \propto
\left(\tilde h_\times  \cos\theta\sin2\varphi 
 - \tilde h_+  \frac{1+\cos^2\theta}{2} \cos2\varphi\right)
$; that of a Michelson rotated by $45^\circ$ from this one is  
$\left(\tilde h_\times \frac{1+\cos^2\theta}{2} \sin 2\varphi
+\tilde h_+\cos\theta\cos 2\varphi \right)$.
 }
\begin{eqnarray}
\left[\tilde{\phi}^{\rm gw}_{\rm DFI}\right]_{\frac{\Omega L}{c} \ll 1}\!\!\!&=&\!\!\! 
{8\sqrt{2}/3 (\Omega L/c)^2 (\omega L /c)}\nonumber \\
\!\!\!&& \!\!\!\!\!\!\!\!\!\!\left[
\tilde h_\times \frac{1+\cos^2\theta}{2} \cos2\varphi
+\tilde h_+\cos\theta\sin2\varphi \right]\!.\quad\;
\end{eqnarray}
For general frequencies and generic incoming GW, the analytical formula for the transfer function is very complicated. Instead, as in Ref.~\cite{KC2}, we show the root-mean-square response function, averaged over GW propagation direction and polarization angle, in Fig.~\ref{fig:resp}. 

\begin{figure}[t]
\includegraphics[width=0.4\textwidth]{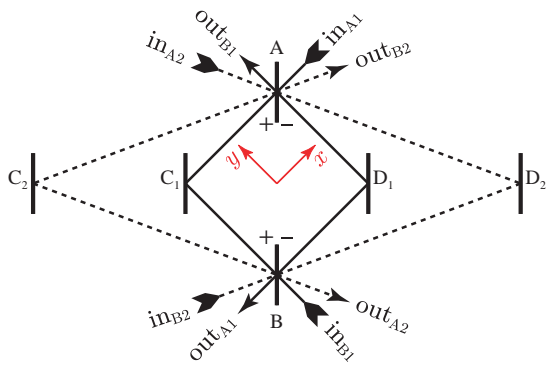}
\caption{The two-dimensional 4-Mach-Zehnder configuration.  \label{2dMZ}}
\end{figure}

{\it 2-D Configuration.} For experimental tests, it is desirable to have a 2-D configuration.  It is straightforward to ``squash'' our 3-D configuration in Fig.~\ref{3dMZ} into a 2-D configuration, as shown in Fig.~\ref{2dMZ}. It also consists of 4 Mach-Zehnder interferometers, $A_1$, $A_2$ (inner Mach-Zehnders, shown in solid lines in the figure), $B_1$, and $B_2$ (outer Mach-Zehnders, shown in dashed lines in the figure). Similar to the 3-D configuration, the subtraction of $B_1$ from $A_1$ cancels displacements of $C_1$ and $D_1$; subtraction of $B_2$ from $A_2$ cancels displacements of $C_2$ and $D_2$. The combination of all four interferometers can cancel motions of the beamsplitters in addition.

It is easy to demonstrate the possibility of canceling beamsplitter displacements without eliminating sensitivity to GWs, by looking at a special case, in which $A C_1 B D_1$ form a square, and 
\begin{eqnarray}
|AC_1|=|AD_1|=|BC_1|=|BC_1|&=&L\,, \\
|AC_2|=|AD_2|=|BC_2|=|BC_2|&=&2L\,.
\end{eqnarray}
We also assume that GW with wavelength $\lambda_{\rm GW}=2L$ propagates perpendicular to the detector plane, with polarization of $\left[\mathbf{e}_{x} \otimes \mathbf{e}_{x} - \mathbf{e}_{y} \otimes \mathbf{e}_{y}\right]$ (see Fig.~\ref{2dMZ}). For the the inner Mach-Zehnders, each of which consists of 4 beams, it is easy to demonstrate that, among them, we have
\begin{eqnarray}
&&\phi_{AC_1}^{\rm gw}=\phi_{C_1B}^{\rm gw}=-\phi_{AD_1}^{\rm gw}=-\phi_{D_2B}^{\rm gw}\,, \nonumber \\
&=&\phi_{BD_1}^{\rm gw}=\phi_{D_1A}^{\rm gw}=-\phi_{BC_1}^{\rm gw}=-\phi_{C_1B}^{\rm gw}\neq0\,.
\label{2dgwphases}
\end{eqnarray}
Here we need to use the fact that GW phase shift flips sign: (i) between beams along $x$ and those along $y$, and (ii) after a time delay of $\lambda_{\rm GW}/(2c)$. For example,
with respect to $\phi_{AC_1}^{\rm gw}$, $\phi_{C_1B}^{\rm gw}$ gains a minus sign twice: the first due to GW polarization, because $AC_1$ is along $-x$, while $C_1B$ is along $-y$; the second because the beam $C_1B$ starts accumulating GW phase shift exactly half an oscillation period after $AC_1$.  Following Eq.~\eqref{2dgwphases}, GW phase gained by all beams in the inner Mach-Zehnders add up, and there is non-vanishing GW response.  On the other hand, the outer Mach-Zehnders do not sense the GW, because GW phase shift, after accumulation by exactly one oscillation period, is zero in each link. This means we can subtract just the correct amount of output from the outer Mach-Zehnders to cancel sensitivity to beamsplitter motions, while keeping the non-vanishing GW response of the inner Mach-Zehnders.  However, as further calculations indicate, the low-frequency response of this 2-D configuration is $\sim f^3$~\cite{KC4}.

{\it Concluding Remarks.} This paper brings Displacement-noise-free Interferometry (DFI) from conceptual plausibility~\cite{KC1,KC2} to concrete and practical optical designs.  We provided simple interferometer configurations that realize DFI.  Compared with the conceptual design in Ref.~\cite{KC2}, our  4-Mach-Zehnder configurations are far more straightforward to implement: instead of requiring composite mirrors with fixed relative positions, our 4-Mach-Zehnder interferometers only require that centers of the multiple beams that reflect off the same mirror must coincide with each other. Moreover, our 3-D configuration has superior low-frequency response ($\sim f^2$, which cannot be exceeded by any DFI configurations) compared to 2-D configurations (which cannot exceed $\sim f^3$).

At the end of this paper, we raise the possibility that our 4-Mach-Zehnder configuration be applied to atomic interferometers proposed for GW detection~\cite{Chiao}, mainly for two reasons: (i) the proposed atomic interferometers already have Mach-Zehnder configurations, and (ii) with much shorter arms compared to long-baseline laser interferometers, displacement noise is likely to become a challenging issue for these detectors.

{\it Acknowledgment.} We thank Curt Cutler for brining Octahedron into our attention. 
Research of Y.C., A.P.\ and K.S.\ are supported by Alexander von Humboldt Foundation's Sofja Kovalevskaja Programme (Funded by the German Ministry of Education and Research). R.L.W.\ was supported by the U.S.\ National Science Foundation under Cooperative Agreement PHY-0107417.

\end{document}